# An approach based on the geometric mean of basic quantitative and qualitative bibliometric indicators to evaluate and analyse the research performance of countries and institutions

Domingo Docampo [1] and Jean-Jacques Bessoule [2,*]

**Short Title : New avenues for the analysis and evaluation of the Research Output of countries and institutions.**

[1] University of Vigo, atlanTTic Research Center for Communications Technologies, Vigo, Spain

[2] CNRS-Univ. Bordeaux, Laboratoire de Biogenèse Membranaire, UMR 5200, Villenave d'Ornon, France

\* Corresponding author: J.-J. Bessoule, UMR 5200 CNRS – Univ. Bordeaux, Bâtiment A3, INRA Bordeaux Aquitaine , 71 Avenue Edouard Bourlaux, CS 20032, 33140 Villenave d'Ornon

e-mail address : jean-jacques.bessoule@u-bordeaux.fr; Tel: (33) 5 57 12 25 70

**Abstract**

We present a straightforward procedure to evaluate the scientific contribution of territories and institutions that combines the size-dependent geometric mean, Q, of the number of research documents (N) and citations (C), and a scale-free measure of quality, q=C/N. We introduce a Global Research Output (GRO-index) as the geometric mean of Q and q. We show that the GRO-index correlates with the h-index, but appears to be more strongly correlated with other well known, widely used bibliometric indicators. We also compute relative GRO-indexes (GROr) associated with the scientific production within research fields. We note that although total sums of GROr values are larger than the GRO-index, due to the non-linearity in the computation of the geometric means, both counts are nevertheless highly correlated. That enables us to make useful comparative analyses among territories and institutions. Furthermore, to identify strengths and weaknesses of a given country or institution, we compute a Relative Research Output count (RROr-index) to tackle variations of the C/N ratio across research fields. Moreover, by using a wealth-index also based on quantitative and qualitative variables, we show that the GRO and RRO indexes are highly correlated with the wealth of the countries and the states of the USA. Given the simplicity of the procedures introduced in this paper and the fact that their results are easily understandable by non-specialists, we believe they could become as useful for the assessment of the research output of countries and institutions as the impact factor is for journals or the h-index for individuals.

**Introduction**

For several years, many powerful indicators have been suggested to evaluate individual research production (Schreiber et al., 2012) and Wildgaard et al. (2014) reviewed 108 of them. Arguably, the h-index (Hirsh, 2005) and the g-index (Egghe, 2006) are the most widely used today to assess individual scientific productions. Several indicators also exist for journals, such as the well-known Impact factor and Scimago Journal Rank (Leydesdorff, 2009). Still, a fair research question of interest for academics, policymakers and the public at large, is how to evaluate the research output of a country or institution. At these macro-levels, the widely used indicators are those found in databases such as Web of Science/Incites: the number of outputs, the number of citations per publication, the number of papers published in the 25% of journals with the highest impact factor for a given research field (Q1), the number of papers in the Top-10% of the most cited papers for a given research field (Top-10), or HCP, the number of highly cited papers (citation thresholds being based on the distribution of citations, picking the specified top fraction of papers for each year and field)… However, a quick look at the list of countries and territories worldwide, along with the data we can gather from the Web of Science shows that evaluating the research output is not such a straightforward task. Take for instance the case of the Belize which, among 189 countries/territories analysed in this paper, would be ranked 144[th] if only the number (N) of Web of Science documents were taken into account, but would be ranked 1[st] in the ratio citations/paper (C/N). It is widely accepted in the bibliometric community that publication and citation measures refer to quantifiable features of research performance in a statistically reliable manner when sufficiently large and preferably longitudinal data sets are available for analysis (Glanzel et al., 2016). And it is quite apparent that the special case of Belize is connected with the scarcity of publications that prevents from the smoothing of ratios, thus resulting in statistically unlikely data. Our goal is not to solve the conundrum of the quantity-quality debate, but rather to help in making comparisons and associations when the datasets under analysis are sufficiently large, *e.g.* comparing the research outputs of France (N = 808198 documents, C/N = 17.75) and Japan (N = 962931, C/N = 13.48), which have a different respective rank if N or C/N is considered. Hence, we aimed to define a bibliometric indicator at macro-levels that combines size-dependent measures, such as the number of documents or citations, with size-independent parameters usually based upon ratios between the number of citations and documents.

In the present paper, we propose an indicator (GRO-index) to evaluate a global research output of countries and institutions that is computed as the geometric mean of a quantitative and a qualitative parameter. Cabrerizo et al., (2010) had also proposed an index based on the geometric average of a quantitative (the h-index) and a qualitative parameter (the m-index) for bibliometric studies at a micro (*i.e.* individual) level. These authors underlined the fact that the use of geometric averages displays several advantages: "it is easy to compute, it is easily understandable in geometric terms, it is not influenced by extremely higher values, and thus, it obtains a value which fuses the information provided by the aggregated values in a more balanced way than other aggregation operators" (Cabrerizo et al., 2010). An indicator based upon the product of a quantitative and a qualitative parameter was also introduced many years ago by Lindsey to assess the scientific production of individual researchers in the Social Sciences (Lindsey 1976, 1978). Glanzel and Moed (2002) pointed to the lack of interpretability in Lindsey's approach as one reason why this indicator has found no application. However, in spite of the fact that his papers were not widely cited, it must be noted that Lindsey's indicator was shown to be useful to make the h-index sensitive to hyper-cited articles (Tahira et al., 2014), and to improve cluster analysis of citation history (Luzar et al., 1992). In 2010, Prathap revisited the Linsey's indicator defined as $CQ = (C^3/N)^{1/2}$. For Prathap, "every citation is actually a paper that cites the publication and has the same dimensions as the h-index (or N)". "Thus, the total received citations C which sums over N has the dimension of area *i.e.*, $h^2$ or $N^2$". And since "CQ does not have the dimensionality of h", Prathap brought it back to the dimensionality of h by introducing a transformation leading to the p-index= $(C^2/N)^{1/3}$ (Prathap, 2010). The weakness of such an approach is that a sum of a sum may not, by and large, lead to the addition of another dimension. As a matter of fact, the number of citations, taken as a sum of a sum of papers, does not behave as a two-dimensional quantity. As N grows (aggregating authors in laboratories, institutions, countries, and/or regions), the number of citations accrued to those N publications remains O(N), since the quotient C/N fluctuates asymptotically around a flat line. In agreement, in the present study, the number of citations for Top-189 countries appears linearly correlated to the number of publications (Figure 1).

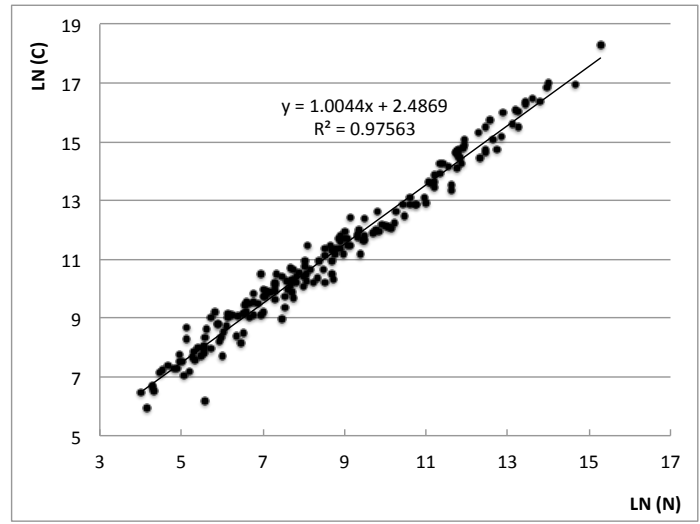

**Figure 1.** Total number of citations as a function of the total number of publications for the Top-189 countries.

Similarly in Prathap (2010), it appears that, at least for Western countries, the C/N ratio did not greatly vary as a function of N. In addition, in the same database used in Prathap (2010), namely https://www.scimagojr.com/countryrank.php, countries as China for example no longer appeared as outliers in 2015, and it can be shown that the number of citations C received by the N papers published in 2015 by the Top-189 countries (*i.e.* countries with N $\geq$ 20, excluding Gibraltar and French territories: Guadeloupe, Martinique, French Polynesia and French Guiana) is C = 5.94N-1375 with $R^2$ =0.977 (data not shown).

The rationale behind the GRO-index we propose in this paper is the following. In the case of a product sold in the market, we compute the total value (V) of the production by multiplying the number of units produced (quantity) by their price (quality). By analogy, an indicator of the value of the research output of a country or institution could be based upon quantitative and qualitative parameters related with their bibliometric output. Moreover, following up on the marketplace analogy, suppose that company A produces high-quality bikes and sells them at twice the worldwide average selling price of bikes. Another company, B, produces low-quality cars and sells them at half the worldwide average selling price of cars. Because of the relative prices of bikes and cars, for the same number of units sold, company A will have a lower market value of production than B, in spite of the higher quality of the units produced by A. Similarly, the most commonly used qualitative measure of the research output of countries, C/N, may greatly vary depending on the research field; it would, therefore, be very useful to introduce as a field-related quality parameter the ratio of institutional or country C/N values to world C/N measures by research field. That qualitative parameter would help us in defining a Relative Research Output Index (RROr-index) per research field.

In this paper, we firstly compare the behaviour of GRO with the usual bibliometric indicators provided by Web of Science/Incites, as well as with the p-index. Some advantages of the GRO-index are further highlighted, then the names and the ranks of Top-34 institutions and TOP-56 countries according to their GRO-values are shown. It is also evidenced that, for both countries and institutions, the sum of the GRO and RRO-indexes computed research field by research field are very efficient linear predictors of the GRO and RRO indexes when all fields are included. This enabled us to analyse the research output of the Top-56 countries research field by research field. Additionally, Abramo and D'Angelo (2016) warned about taking size-independent citation indicators per se as indicators of research performance, unless they are placed in context through accompanying measures of expenditures on research and or researchers. The use of ratios may be shown to violate basic economic reasoning accepted facts such as "that the better performer under parity of resources is the actor who produces more; or under parity of output, the better is the one who uses fewer resources" (Abramo and D'Angelo, 2016). Hence, we would also like to connect bibliometric information with appropriate measures of expenditures or wealth, to shed light on the relative performance and efficiency of countries and territories.

**Material and Methods**

Raw bibliometric data for our analysis were extracted from the InCites platform, provided by Clarivate Analytics Philadelphia, Pennsylvania, USA (2017), under the Essential Science Indicators scheme, including "articles and reviews from Science Citation Index Expanded and Social Science Citation Index". The dataset does not include "publications from Arts & Humanities, Conference Proceedings Citation Index, or Book Citation Index". Our raw data then comprises articles and reviews published between 2006 and 2015. Only countries showing more than 50 documents in the period were analysed. In addition, England, Northern Ireland, Wales and Scotland were removed from the analyses to avoid redundancy with the data from the United Kingdom. For the same reasons, results for Netherlands Antilles and French territories (New Caledonia, Reunion, French Guiana, French Polynesia, Guadeloupe, Martinique) were not taken into account. Nevertheless, it must be noted that following all analyses, these 11 entities did never appear as outliers, and results would be extremely similar if considered. Subsequently to the treatment of our data, a total of 189 countries were analysed. We also analysed up to 4556 Institutions: 2965 so-called "Academics" (mainly universities), 890 "Research Institutes", 331 "Health", 186 "Corporates", 136 "Governments", plus 48 institutions belonging to various other institutional categories ("Laboratory", "Museum", "Observatory"…). To avoid redundancy, the 76 University Systems were not taken into account.

Financial data were gathered from the International Monetary Fund's World Economic Outlook (IMF, 2017) on 2018, January $4^{th}$. Gross domestic products at current prices (GDP values) are based upon GDP in national currency converted to U.S. dollars using market exchange rates (yearly average). For the present analysis, the variable GDP corresponds to the average of the values of GDP along the period 2006-2015. Gross domestic product based on purchasing-power-parity per capita, GDP at current international dollars (PPC) were expressed in GDP in PPP dollars per person. Data are derived by dividing GDP in PPP dollars by total population. For the present analysis, the variable PPC corresponds to the average of the values of GDP (PPP) per capita along the period 2006-2015. From the two financial variables collected, a wealth index, WTH, has been composed as the geometric mean of GDP and PPC.

The (ISO alpha-3) three letter codes were used to designate countries: ARG: Argentina, AUS: Australia, AUT: Austria, BEL: Belgium, BGR: Bulgaria, BRA: Brazil, CAN: Canada, CHE: Switzerland, CHL: Chile, CHN: China, COL: Colombia, CZE: Czech Republic, DEU: Germany, DNK: Denmark, EGY: Egypt, ESP: Spain, EST: Estonia, FIN: Finland, FRA: France, GBR: United Kingdom, Greece: GRC, HKG: Hong Kong, HRV: Croatia, HUN: Hungary, IND: India, IRL: Ireland, IRN: Iran, ISL: Iceland, ISR: Israel, ITA: Italy, JPN: Japan, KEN: Kenya, KOR: Korea, MEX: Mexico, MYS: Malaysia, NLD: Netherlands, NOR: Norway, NZL: New Zealand, PAK: Pakistan, POL: Poland, PRT: Portugal, ROU: Romania, RUS: Russia, SAU: Saudi Arabia, SGP: Singapore, SRB: Serbia, SVK: Slovakia, SVN: Slovenia, SWE: Sweden, THA: Thailand, TUR: Turkey, TWN: Taiwan, UKR: Ukraine, USA: United States of America, ZAF: South Africa.

**Rationale**

To evaluate the research output of a country or institution, we consider two quantitative parameters, namely N, number of articles or reviews published between 2006 and 2015, and C, number of citations accrued to those papers between 2006 and 2017. A plot of C *vs* N in a double logarithmic scale is shown in Figure 1. Since the value of the slope in Figure 1 is very close to 1, C and N appear linearly correlated. Therefore any of these parameters can be *a priori* chosen as the quantitative parameter to evaluate a research output. Nevertheless, to not favour one parameter over the other, the geometric mean of N and C will be considered as the quantitative parameter of the scientific production of a country or an institution: $Q = (N.C)^{1/2}$. The second parameter we use to evaluate the Global Research Output (GRO) of a country or institution is a qualitative measure, *i.e.* number of citations per publication, $q = C/N$. Now, to combine the quantitative and qualitative measures, Q and q, we take the geometric mean of both parameters, producing the GRO-index = $(Q.q)^{1/2} = (C^3/N)^{1/4}$. The p-index can be also regarded as the geometric mean between a quantitative parameter ($C^{2/3}$) and a qualitative one $(C/N)^{2/3}$. But GRO appears more balanced since it aggregates both the number of publications and the number of citations within the quantitative parameter, thus giving equal importance to both C and N in the computation of the index. As an example, the calculations for the world all fields included ($GRO_w$), as well as research field by research field ($GROr_w$), are shown in Table 1.

**Table 1**: Raw Data and Calculation of the GRO-index and the various GROr for the world.

|  | $N_w$ | $C_w$ | $Q_w$ | $q_w$ | $GRO_w$ |
|---|---|---|---|---|---|
| **All Fields** | 12669278 | 213945356 | 52062781 | 16.89 | 29651 |
| **Research Fields** | $Nr_w$ | $Cr_w$ | $Qr_w$ | $q_{rw}$ | $GROr_w$ |
| Agricultural Sciences | 350182 | 4214897 | 1214900 | 12.04 | 3824 |
| Biology & Biochemistry | 655603 | 15040990 | 3140210 | 22.94 | 8488 |
| Chemistry | 1489725 | 28295481 | 6492495 | 18.99 | 11105 |
| Clinical Medicine | 2350035 | 40702524 | 9780202 | 17.32 | 13015 |
| Computer Science | 304964 | 3505363 | 1033929 | 11.49 | 3447 |
| Economics & Business | 229392 | 2820097 | 804306 | 12.29 | 3145 |
| Engineering | 1006427 | 11974992 | 3471593 | 11.90 | 6427 |
| Environment/Ecology | 370443 | 6883427 | 1596846 | 18.58 | 5447 |
| Geosciences | 377291 | 6472810 | 1562733 | 17.16 | 5178 |
| Immunology | 224428 | 5737206 | 1134720 | 25.56 | 5386 |
| Materials Science | 658410 | 10850762 | 2672873 | 16.48 | 6637 |
| Mathematics | 370480 | 2267589 | 916568 | 6.12 | 2369 |
| Microbiology | 177598 | 3687959 | 809305 | 20.77 | 4099 |
| Molecular Biol. & Genetics | 394274 | 13354013 | 2294589 | 33.87 | 8816 |
| Multidisciplinary | 16759 | 355231 | 77158 | 21.20 | 1279 |
| Neuroscience & Behavior | 452541 | 11128242 | 2244100 | 24.59 | 7429 |
| Pharmaco. & Toxicology | 340426 | 5950164 | 1423232 | 17.48 | 4988 |
| Physics | 1024499 | 15784117 | 4021295 | 15.41 | 7871 |
| Plant & Animal Science | 644911 | 8171135 | 2295573 | 12.67 | 5393 |
| Psychiatry/Psychology | 339808 | 5959524 | 1423058 | 17.54 | 4996 |
| Social Sciences, general | 758582 | 7542918 | 2392054 | 9.94 | 4877 |
| Space Science | 132500 | 3245915 | 655808 | 24.50 | 4008 |

From Table 1, it appears that the qualitative parameter $q_{rw}$ greatly varies from 6.12 (Mathematics) to 33.87 (Molecular Biology & Genetic). Hence, to analyse strengths and weaknesses within the same country or institution across research fields, we also calculated a RROr-index by research field. It is calculated as the GROr-index for the research field r, but using as the qualitative parameter q: the $Cr_x/Nr_x$ ratio for the country or institution x, divided by the world $Cr_w/Nr_w$ ratio for the same research field. A RRO-index all fields included can be also calculated, and it can be shown that $RRO = (q_w)^{-1/2}.GRO$.

**Results and Discussion**

In contrast with countries for which, to our knowledge, no h-indexes were available on the INCITES platform, the study of Institutions provided the opportunity to analyse correlations between the GRO- and h- indexes. By taking into account all of the 4556 Institutions analysed, it is apparent that the GRO-and h- indexes are highly correlated ($R^2$=0.980, not shown). If only the 1205 Institutions displaying an h-index higher than 100 were taken into account, the high correlation still holds ($R^2$=0.977, not shown). Even focusing on the 129 Institutions with a GRO-index higher than 2000, the correlation is still very significant ($R^2$=0.932, Figure 2).

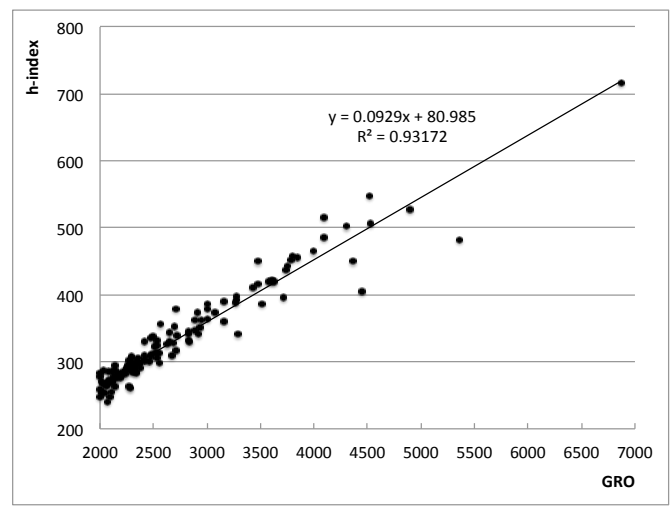

**Figure 2:** h-index as a function of GRO-index for Institutions displaying a GRO-index higher than 2000.

While h- and GRO-indexes of institutions displayed high correlation values, yet the GRO-index exhibits advantages vis-à-vis the h-index in order to analyse the scientific production of an institution or a country, namely:

(i) Unlike the GRO-index, a given h-index would not increase when new publications are added to the sample, unless the number of their citations exceeds the h-value;

(ii) The GRO-index is a much simpler and therefore transparent tool than the h-index considering its formula: square root [C/N. square root (C.N)]. Its limpidity enables its use to calculate scientific production of institutions and countries over several years, differentiating itself from the complexity of the h-index;

(iii) Just by adding the number of publications and by adding the number of citations, it is easier to calculate the GRO-index than the h-index of a group of countries or Institutions. The same remark can be made for the analysis of a group of several research fields (e.g. the various research fields addressing Human and Social Sciences);

(iv) Thanks to the "market value of industry production" analogy, the GRO-index can be easily explained and understood. This is a critical point, especially to reach a wider audience of non-specialists, including policy makers. In contrast, it is more complex to explain why the h-index is such a valuable indicator;

(v) As shown in Figure 3, in comparison with the h-index, the GRO-index appears to be more strongly correlated with other indicators such as Q1, Top-10, and HCP. If we considered the p-index, it appeared that it is even less correlated to these indicators than the h-index. It must be noted that in contrast with the p-, h- and GRO- indexes, Q1, TOP 10% and HCP greatly depend on the amount and quality of the scientific production of other countries/Institutions.

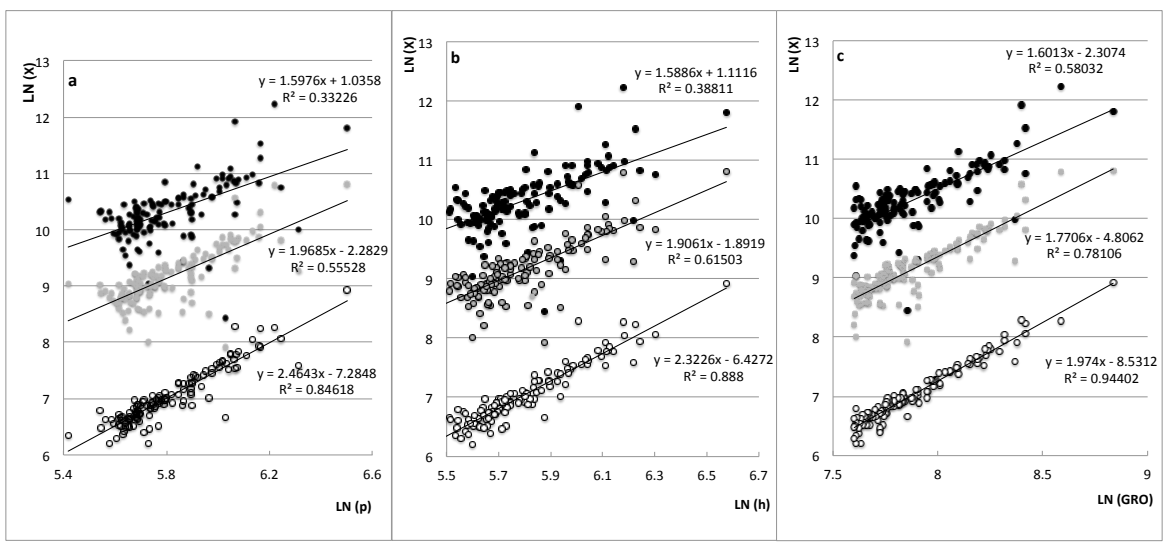

**Figure 3:** LN(Q1), LN(Top-10), and LN(HCP) as a function of LN(p), LN(h) and LN(GRO) for Institutions displaying a GRO-index higher than 2000.

a: LN(X) as a function of LN(p); b: LN(X) as a function of LN(h); c: LN(X) as a function of LN(GRO). X = Q1 (closed circles), Top-10 (grey circles) or HCP (open circles).

Hence, GRO appears as a global and efficient indicator to evaluate Institutions. The names of Institutions with a GRO-index higher than 3000, are indicated in Table 2.

Table 2 Rank, Name, GRO-index values of the Top-34 Institutions.

| Rank | Name | GRO |
|---|---|---|
| 1 | Harvard University | 6875 |
| 2 | Centre National de la Recherche Scientifique (CNRS) | 5353 |
| 3 | National Institutes of Health (NIH) - USA | 4902 |
| 4 | United States Department of Energy (DOE) | 4525 |
| 5 | VA Boston Healthcare System | 4522 |
| 6 | Chinese Academy of Sciences | 4444 |
| 7 | Max Planck Society | 4360 |
| 8 | Howard Hughes Medical Institute | 4306 |
| 9 | Stanford University | 4096 |
| 10 | Massachusetts Institute of Technology (MIT) | 4088 |
| 11 | Johns Hopkins University | 3994 |
| 12 | University of Toronto | 3841 |
| 13 | University of California Berkeley | 3794 |
| 14 | University of Washington Seattle | 3780 |
| 15 | University of California Los Angeles | 3753 |
| 16 | University of Oxford | 3735 |
| 17 | Institut National de la Sante et de la Recherche Medicale (Inserm) | 3712 |
| 18 | University of Michigan | 3634 |
| 19 | University of California San Francisco | 3606 |
| 20 | University of Pennsylvania | 3606 |
| 21 | University of Cambridge | 3574 |
| 22 | University College London | 3514 |
| 23 | Massachusetts General Hospital | 3482 |
| 24 | Columbia University | 3473 |
| 25 | University of California San Diego | 3434 |
| 26 | Consejo Superior de Investigaciones Cientificas (CSIC) | 3289 |
| 27 | Duke University | 3278 |
| 28 | Yale University | 3276 |
| 29 | Imperial College London | 3260 |
| 30 | University of Chicago | 3162 |
| 31 | Pierre & Marie Curie University - Paris VI | 3156 |
| 32 | Cornell University | 3078 |
| 33 | University of Pittsburgh | 3005 |
| 34 | Washington University (WUSTL) | 3003 |

Similarly, Figure 4 shows the GRO-index of all the countries (log scale) ordered from highest to lowest. As it can be seen in Figure 4, excluding the highest performer (USA), the Top-189 countries can be classified into four groups: 16 countries with a GRO-index higher than 4000 and ranked between 2 and 17; 38 countries ranked between 18 and 55 with a GRO-index ranging from 3500 to approx. 1000; 99 countries ranked between 56 and 114 and displaying GRO-indexes between 850 and 150; plus 35 countries with the lowest GRO-indexes.

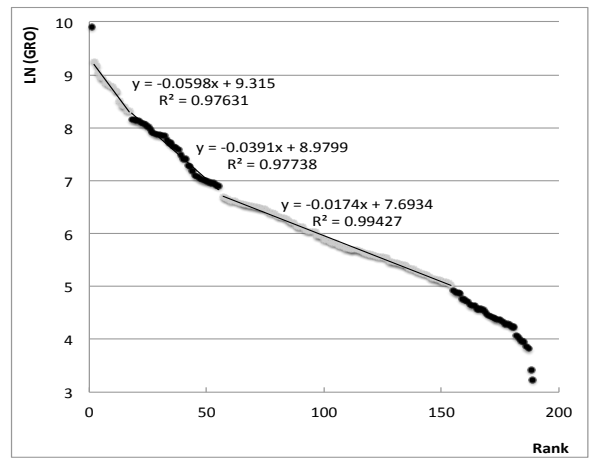

**Figure 4:** LN(GRO) vs GRO-Rank for the Top-189 countries.

The name and the GRO-index of the Top-56 countries are shown in Table 3.

In addition to GRO and RRO, GROr and RROr indexes can be used to highlight specific strengths within a given institution or country by calculating them for a given research field r. But the sums of the RROr and GROr indexes obtained for all the Research Fields are not equal to RRO and GRO indexes due to the non-linearity in the computation of the indexes. For example $GRO_{world}$ calculated in the last line of the Table 1 is $0.231 \cdot sum(GROr_{world})$. To add up all the GROr field indexes together makes sense, since there is no overlapping among the 22 ESI research fields. The relevant question is whether that sum of the 22 ESI fields constitutes a reasonable approximation for the global GRO-index. To explore the answer to this important question we have collected values from Institutions for which data were available. Institutions were firstly clustered into two groups: 2766 "Academic" institutions (mainly corresponding to universities) and 1475 "Other" institutions. For Academic institutions a strong ($R^2 = 0.983$) linear relationship between GRO and sum (GROr) was observed, and the slope GROr/sum(GROr) = 0.249. For other institutions, the correlation coefficient is lower: $R^2 = 0.938$, and the slope is higher (0.289). We further investigated the reasons for these differences.

**Table 3** Rank, GRO-index of theTop-56 countries and their WTH values.

| Rank | Country | GRO | WTH | Rank | Country | GRO | WTH |
|---|---|---|---|---|---|---|---|
| 1 | USA | 20049 | 900.1 | 29 | RUS | 2634 | 199.5 |
| 2 | GBR | 10417 | 321.9 | 30 | IRL | 2618 | 112.7 |
| 3 | DEU | 9581 | 386.8 | 31 | TUR | 2584 | 124.4 |
| 4 | CHN | 8377 | 277.4 | 32 | NZL | 2562 | 71.9 |
| 5 | FRA | 7773 | 321.3 | 33 | IRN | 2373 | 82.2 |
| 6 | CAN | 7576 | 259.5 | 34 | ZAF | 2239 | 63.9 |
| 7 | JPN | 6903 | 434.8 | 35 | CZE | 2234 | 75.8 |
| 8 | ITA | 6874 | 273.7 | 36 | MEX | 2099 | 136.2 |
| 9 | NLD | 6555 | 196.7 | 37 | ARG | 2038 | 94.2 |
| 10 | AUS | 6370 | 232.1 | 38 | HUN | 1979 | 55.8 |
| 11 | ESP | 6081 | 214.3 | 39 | CHL | 1767 | 67.7 |
| 12 | CHE | 5878 | 182.0 | 40 | THA | 1650 | 69.4 |
| 13 | SWE | 4937 | 148.7 | 41 | SAU | 1632 | 170.2 |
| 14 | KOR | 4587 | 192.8 | 42 | EGY | 1452 | 51.2 |
| 15 | BEL | 4484 | 141.4 | 43 | MYS | 1414 | 76.8 |
| 16 | DNK | 4144 | 119.1 | 44 | SVN | 1322 | 37.1 |
| 17 | IND | 4115 | 91.0 | 45 | PAK | 1215 | 30.1 |
| 18 | AUT | 3500 | 132.8 | 46 | ROU | 1197 | 56.4 |
| 19 | BRA | 3472 | 169.2 | 47 | ISL | 1138 | 26.2 |
| 20 | TWN | 3419 | 134.5 | 48 | COL | 1121 | 58.3 |
| 21 | ISR | 3392 | 87.0 | 49 | EST | 1099 | 23.7 |
| 22 | SGP | 3250 | 135.7 | 50 | HRV | 1092 | 34.4 |
| 23 | FIN | 3201 | 101.0 | 51 | SVK | 1072 | 47.4 |
| 24 | NOR | 3144 | 167.9 | 52 | SRB | 1043 | 22.6 |
| 25 | HKG | 3011 | 112.0 | 53 | UKR | 1042 | 33.9 |
| 26 | POL | 2843 | 103.2 | 54 | KEN | 1003 | 11.1 |
| 27 | GRC | 2719 | 87.1 | 55 | BGR | 986 | 29.1 |
| 28 | PRT | 2675 | 77.9 | 56 | TUN | 845 | 21.2 |

By considering institutions focused in one major research field, the more specialised the institution becomes, the higher its max(GROr/sum(GROr)) is. We thus took SGr = max(GROr/sum(GROr)) as a specialization index of an institution. We analysed the variation of the GROr/sum(GROr) ratio as a function of SGr. It appears that the GRO/sum(GROr) ratio increases with SGr and that the slope of the straight line is lower for Academic institutions (Figure 5a) than for the "Other" institutions (Figure 5b). Overall, Academic institutions are by and large more comprehensive and thus show much lower specialization indexes, SGr, than institutions classified as "Other". Nevertheless an institution involved in only two research fields with two GROr/sum(GROr) ratio = 0.5 could be regarded as more specialised than an institution with one major research field displaying a highest GROr/sum(GROr) ratio =0.6, and 4 other minor research fields displaying GROr/sum(GROr) ratio =0.1. Thus we also analysed data by considering the sum of the two highest GROr/sum(GROr) ratios of the institutions: the conclusions were the same.

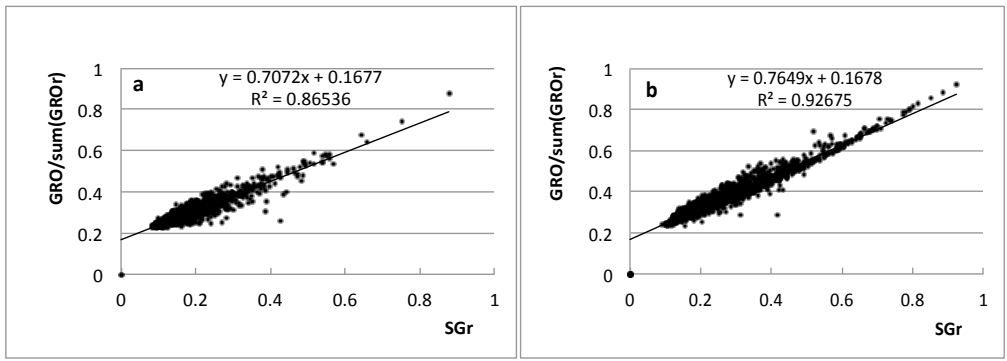

**Figure 5**: GROr/sum(GROr) as a function of SGr for Academic Institutions and "Other" Institutions.
a: Academic Institutions. b: Other Institutions.

Hence it appears that differences in the correlation coefficients and in the slopes of the straight lines GRO = f(sum(GROr)) were observed because Academic institutions are globally less specialised than the "Other" Institutions. In fact, only 5.8% of Academic institutions display a specialisation index, SGr, higher than 0.3 compared to 39.6% for "Other" institutions, while 107 "Other" Institutions (7.3%) display a specialisation index higher than 0.5, compared to 21 for Academic Institutions, *(*0.76%). Moreover, eight of those 21 institutions could be regarded as "Research Institutes", and therefore classified as "Other" institutions. These eight institutions are the following: Institute of Physics of the Azerbaijan National Academy of Sciences; Instituto de Fisica Corpuscular; Yerevan Physics Institute; Institut des Hautes Etudes Scientifiques; Institut d'Optique Graduate School – Dublin; National Research Nuclear University -Moscow Engineering Physics Institute, Dublin Institute for Advanced Studies; National Centre for Physics - Pakistan. Same comment can be made for two other institutions, namely the European Southern Observatory and the Warsaw University Observatory. In addition, it is not surprising to find among these 21 institutions, entities such as Ufa State Aviation Technical University, National Research Nuclear University, Paris School of Economics and Princeton Plasma Physics Laboratory.

Such an analysis of the correlation between GRO and sum(GROr) was repeated for the sample of 189 countries worldwide. In this case, the correlation reaches 0.999 (Figure 6a). The slope of the straight line (0.225) is very close to the $GRO_{world}$/sum($GROr_{world}$) ratio (0.231) calculated from results shown in Table 1. It is also close to the slope determined for Academic institutions (0.249). Interestingly, for the 189 countries, the same correlation coefficient ($R^2$ =0.999) and a similar slope of the straight line (0.219) are observed by plotting RRO as the function of sum(RROr) (Figure 6b). Given that the sums of the GROr and RROr are very efficient linear predictors of the GRO and RRO index respectively, research field scores constitute a share of the index and enable to profile the research output of countries. Moreover RRO/GRO = $(q_w)^{-1/2}$, and since for countries, GRO = a.sum(GROr), and RRO = b.sum(RROr), it appears that any one of these indexes is sufficient to analyse their research output. However, to compare two very specialised institutions focusing on two research fields displaying very different $q_w$ values (as for example Mathematics and Molecular Biology & Genetics, see Table 1), it can be much more sound to use the RRO-index rather than the GRO-index.

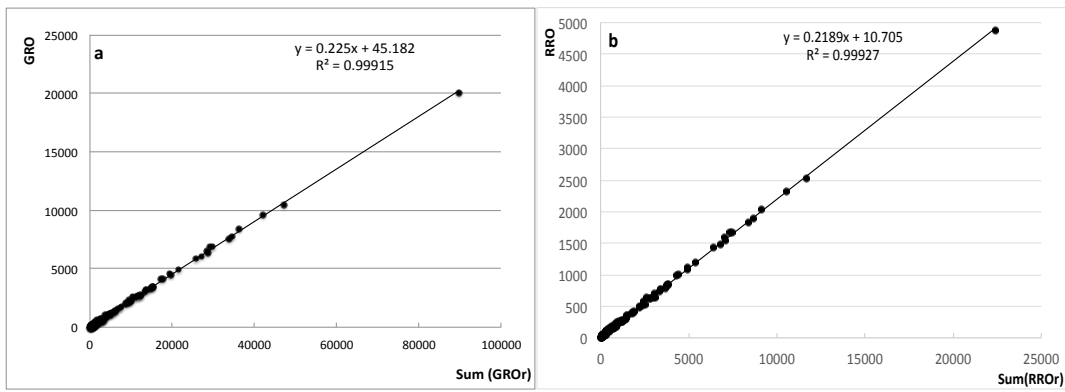

**Figure 6:** GRO as a function of sum(GROr) and RRO as a function of sum(RROr) for 189 countries.

a: GRO as a function of sum(GROr); b: RRO as a function of sum(RROr).

To better understand the composition of the GRO index in terms of its Field-constituents, we carried out an exploratory analysis on the set of GROr values for a large number of academic institutions (close to 2000) worldwide. We found six principal components with an eigenvalue larger than one, which makes it difficult to reduce dimensionality and interpret the results at the same time. To gain more insight into the Field distribution we used a mixed approach. We first carried out a hierarchical cluster analysis over the set of 22 scores on the research fields. We found that the 22 fields can be adequately classified in five clusters, as shown in Table 4, in which the names of the clusters try to describe the areas inside.

**Table 4** Cluster analysis on the 22 research fields' GRO values.

| Cluster | Research Fields |
|---|---|
| agrenv | Agricultural Sciences/ Environment & Ecology/ Plant & Animal Sciences |
| medlife | Biology & Biochemistry/ Molecular Biology & Genetics/ Clinical Medicine/ Microbiology Multidisciplinary/ Immunology /Neurosciences & Behavior/ Pharmacology & Toxicology |
| chemateng | Chemistry/ Materials Science/ Engineering/ Computer Science/ Mathematics |
| socsci | Psychiatry & Psychology/ Social Sciences, general/ Economics & Buisness |
| geophy | Physics/ Space Science/ Geosciences |

Now, because GRO indexes can be aggregated, we computed the scores on the five aggregated sets of research fields according to the cluster compositions. We then analysed the data corresponding to the 56 countries shown in Table 3 and carried a principal component analysis to reduce dimensionality on the new five aggregated variables. We used the covariance matrix since data are commensurable. We found two principal components with eigenvalues in excess of 1.0 that account for more than 70% of the variance of the sample. The scores of the Top-countries in the two components along with the location of the variables are shown in Figure 7.

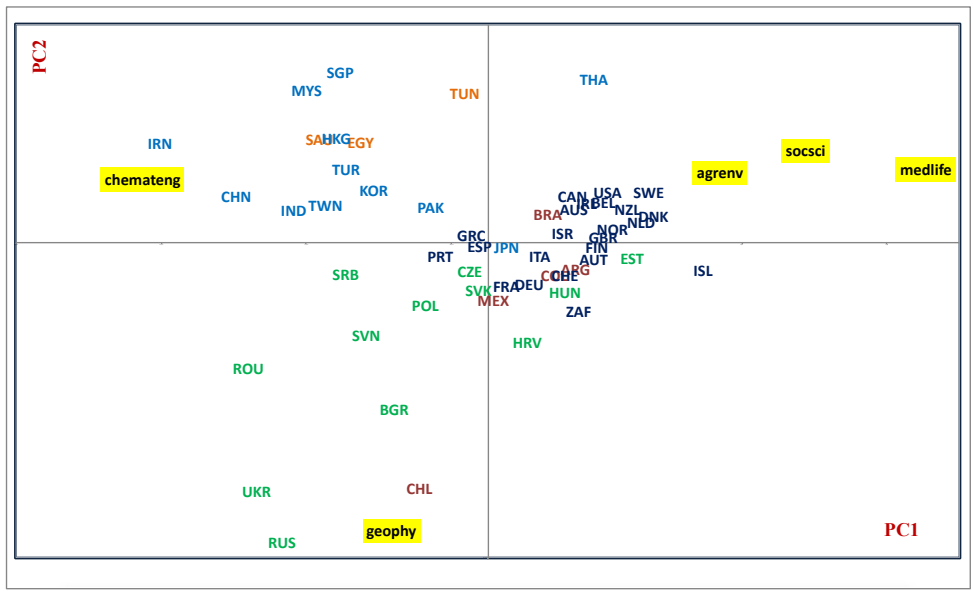

**Figure 7**: Scores on the two principal components (sample of 55 countries). Kenya (outlier) was omitted. Light blue, orange, brown and dark blue letters correspond to Asian, North-African, Latin American, and Western (+ South Africa) Countries respectively. Green letters correspond to countries of former Soviet-Union and satellite countries

Except Japan, all Asian and North-African countries are found above the horizontal axis, and with the exception of Thailand, all of them are in the upper-left quadrant, corresponding to Chemistry, Materials Science, Mathematics, and Engineering. By contrast, all countries of former Soviet-Union and satellite countries are found below the horizontal axis, and with the exception of Hungary, Croatia and Estonia, they are plotted in the lower-left quadrant corresponding to Physics, Space Science, and Geosciences. Western countries, Latin American countries, as well as South Africa were close to the horizontal axis, and rather located at the right side of the vertical axis (Life Sciences plus Human and Social Sciences). Chile, located at the bottom of the lower-left quadrant is an exception. This is due to its strong involvement in Space Sciences: 14% of sum(GROr), in comparison with all other countries (between 0.26% and 7.1%). This can be easily explained by the presence of many high-end astronomical observatories in this country.

We then analysed the relationship between the wealth of countries and their GRO-index. As mentioned in Material and Methods, and as done for the research output of countries, the wealth of a country (WTH) was defined as the geometric mean of a quantitative parameter (GDP: Gross Domestic Product in US$) and a qualitative one (PPC: GDP in PPP US$ per capita), averaged in both cases over the period 2006-2015). The WTH Values of the Top-56 countries are shown in Table 3. With the exception of Kenya (high GRO/WTH ratio) and Saudi Arabia (low GRO/WTH ratio) which constitute apparent outliers, there is clear evidence supporting a strong linear relationship between LN(WTH) and LN(GRO), as Figure 8 shows.

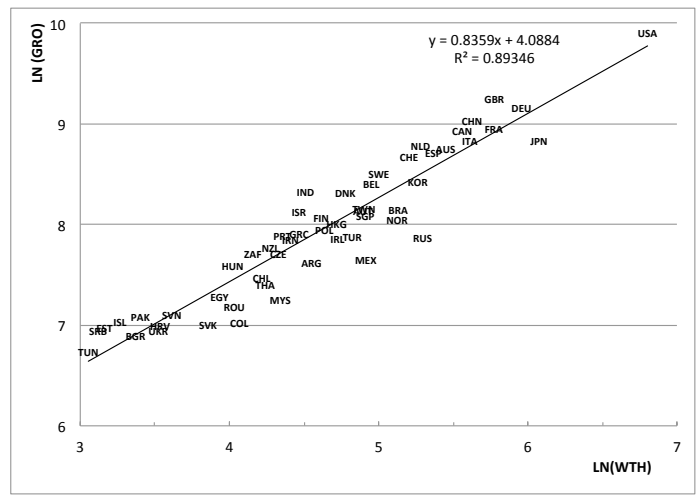

**Figure 8:** LN(GRO) as a function of LN(WTH) for 54 countries among the Top-56. Kenya and Saudi Arabia were omitted.

Interestingly, with the exception of Maryland (home of the National Institute of Health, Howard Hughes Medical Institute, Johns Hopkins University…) and Massachusetts (Massachusetts Institute of Technology, Harvard…) which appeared as outliers, Figure 9 shows that results for the states of the USA were consistent with the ones shown for countries in Figure 8.

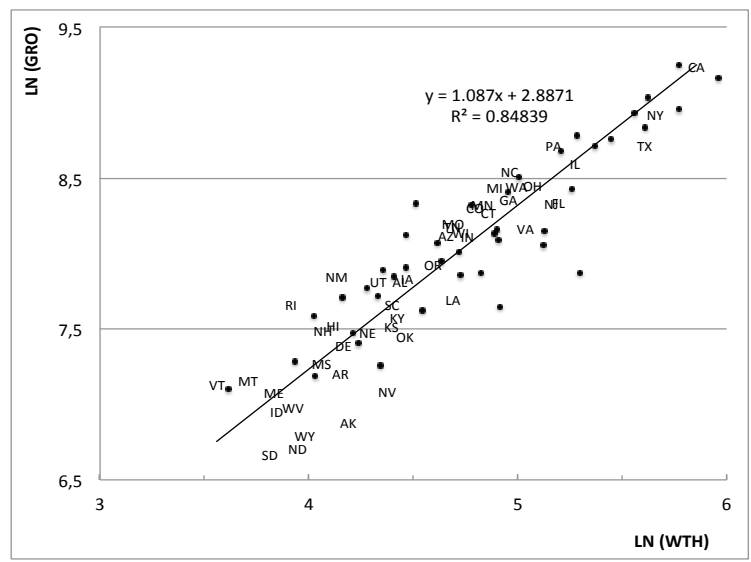

**Figure 9:** LN(GRO) as a function of LN(WTH) for 48 US states. Dots correspond to countries with $6 > LN(WTH) > 3.5$ shown in Figure 8. Maryland and Massachusetts were omitted.

It comes as no surprise that the research output of a country/state is by and large commensurate with its wealth; whether wealth is the scientific progress driver or the other way around is a debate. Wealth and knowledge production appear nowadays so intertwined that it is very difficult to answer the question of which causes which, although the modern endogenous growth theory states that the stock of human capital is an endogenous source of technological change which determines the rate of growth (Romer, 1990), thus solidly linking the production of knowledge with the wealth of nations.

To summarize, the use of the CQ-like indicator introduced in this paper to assess the relative strength of the performance of countries in the 22 research fields into which INCITES splits the bibliometric data makes it possible to evidence that their research output is greatly related to the geographical- historical- and economic -contexts. Reasonably combining quantitative and qualitative data is arguably a matter of great interest to inform decision- and policy-making within institutions or whole research systems.

**Acknowledgments**

We thank Dominique Dunon-Bluteau and Daniel Egret for initiating the human link between the authors. We are grateful to Paul Gouguet, Amélie Bernard and Pierre Madre for critical reading of the manuscript.